# Hydrodynamic drift ratchet scalability

J. W. Herringer [1], D. Lester [2], G. E. Dorrington [1], J. G. Mitchell [3] and G. Rosengarten[1]

[1] *School of Aerospace, Mechanical and Manufacturing Engineering, RMIT University, Carlton, Melbourne, Victoria, 3053, Australia*

[2] *School of Civil, Environmental and Chemical Engineering, RMIT University, Carlton, Melbourne, Victoria, 3053, Australia*

[3] *School of Biological Sciences, Flinders University, Adelaide, South Australia, 5001, Australia*

The rectilinear "drift" of particles in a hydrodynamic drift ratchet arises from a combination of diffusive motion and particle-wall hydrodynamic interactions, and is therefore dependent on particle diffusivity, particle size, the amplitude and frequency of fluid oscillation and pore geometry. Using numerical simulations, we demonstrate that the drift velocity relative to the pore size is constant across different sized drift ratchet pores, if all the relevant non-dimensional groups (Péclet number, Strouhal number and ratio of particle to pore size) remain constant. These results clearly indicate for the first time the scaling parameters under which the drift ratchet achieves dynamic similarity, and so facilitates design, fabrication and testing of drift ratchets for experiments and eventually as commercial micro/nano fluidic separation devices.

a) Electronic mail: j.herringer@student.rmit.edu.au
b) Electronic mail: graham.dorrington@rmit.edu.au
c) Electronic mail: daniel.lester@rmit.edu.au
d) Electronic mail: jim.mitchell@flinders.edu.au
e) Electronic mail: gary.rosengarten@rmit.edu.au

# 1. INTRODUCTION

Drift ratchets have received considerable attention over the past decade due to their intriguing non-equilibrium thermodynamic properties and myriad of potential applications (J. C. T. Eijkel & van den Berg, 2006; J. T. Eijkel & Berg, 2005; Kettner, Reimann, Hänggi, & Müller, 2000). A drift ratchet is a micro/nano fluidic device consisting of oscillating zero-mean fluid flow in a series of periodic ratchet-shaped pores, Fig. 1, which generates rectified motion of finite-sized colloidal Brownian particles (Kettner et al., 2000; Matthias & Muller, 2003).

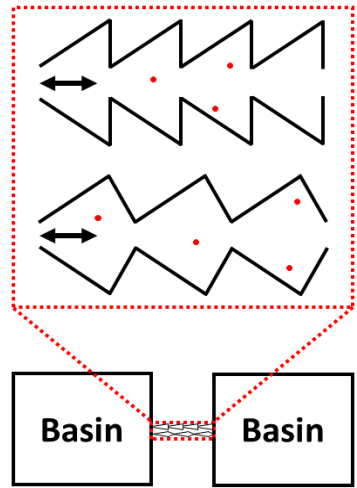

**Fig. 1** Asymmetric, ratchet-shaped pore profiles, with the arrows indicating oscillating fluid motion.

This device operates at nano- to microscopic spatial scales where Brownian motion is a dominant transport mechanism, and converts random thermal motion into directed particle motion. Whilst such disordered diffusion may not be expected to generate mean particle flux due to the Second Law of Thermodynamics, these systems are driven far from thermal equilibrium, and therefore may be considered as open weakly dissipative systems. Rectified particle transport arises from a combination of irreversible Brownian motion and symmetry breaking due to hydrodynamic interactions between the advecting particle and asymmetric ratchet walls. The rectified rate of movement in one direction along the pore axis is commonly termed the drift velocity.

The drift velocity magnitude and direction are dependent upon a combination of the basic particle physical properties (e.g. size and shape) and ratchet geometry. Hence - if designed and tuned correctly - the drift ratchet is capable of continuous particle separation on the basis of small differences in basic particle properties alone. Separation is a critical step in many micro/nano fluidic applications and lab-on-a-chip devices and these drift ratchets offer a unique and fundamentally simple technique for it. Also, if the mechanism of separation is understood properly, it may subsequently lead to further insights into transport in biological systems (Losic, Rosengarten, Mitchell, & Voelcker, 2006; Yang, Lopez, & Rosengarten, 2011). Since the original computational (Kettner et al., 2000) and experimental (Matthias & Muller, 2003) studies of the drift phenomenon over a decade ago, there has been a flurry of research activity (Carbajal-Tinoco, Lopez-Fernandez, & Arauz-Lara, 2007; Klaus, Frank, & Ulrich, 2011; Kondratyev, Urbano, & Vorotnikov, 2014; Martens, Schmid, Schimansky-Geier, & Hänggi, 2011; Perkins & Jones, 1992) regarding these devices and the thermodynamic principles upon which they operate.

Whilst the applications of the drift ratchet are immense, to date these have been largely untapped. The immaturity of this technology is largely due to the lack of efficient quantitative models for the prediction of the drift velocity, the unresolved debate regarding the physical mechanisms upon which the drift ratchet operates, and most importantly the lack of experimental confirmation. Whilst Matthias and Muller (2003) appear to demonstrate action of the ratchet phenomenon, subsequent studies (Klaus et al., 2011) cast some doubt upon these results but do not totally dismiss the existence of the drift ratchet phenomena. In light of these limitations, we have explored the dynamic similarity of drift ratchets to aid experimental design. We have developed a numerical model and completed simulations to both verify the scaling properties of the drift ratchet, and understand the physical mechanisms that generate rectified particle motion.

The initial computational drift ratchet study of Kettner et al. (2000) involved a simplified model of particle-wall hydrodynamic interactions, which, demonstrated that finite particle size is necessary to invoke rectified motion. Later studies (Ai & Liu, 2013; Makhnovskii, Zitserman, & Antipov, 2012; Martens et al., 2011; Reguera et al., 2012) made progress in explaining the drift ratchet mechanism via the Fick-Jacobs approximation based upon entropic arguments which attempts to quantify the augmented particle diffusivity in confined geometries. However, the applicability of this approximation to hydrodynamic problems is currently unresolved (Martens, Schmid, Straube, Schimansky-Geier, & Hänggi, 2013).

The computational results of Kettner et al. (2000) were experimentally replicated at a qualitative level by Matthias and Muller (2003), who observed unidirectional drift of spherical particles in a massively parallel drift ratchet membrane driven by an oscillatory pressure-driven flow. These experiments appeared to verify the numerical results of Kettner et al. (2000) and clarify the influence of the fluid oscillation amplitude upon drift current. This was done by comparing the average drift velocity of particles in simulations with the measured change in concentration of fluorescent micro-particles in the experiments.

More recently, Klaus et al. (2011) attempted to replicate these experiments by Matthias and Muller (2003) - albeit with a slightly different pore geometry – but found drift also occurred in straight-walled cylindrical pores. The authors concluded that the particle transport was due to advection under pressure-driven oscillatory flow, inducing non-zero mean flow rather than a ratchet mechanism. Specifically, they could not confirm that the fluid volume displaced over an oscillation half period was not less than the total pore volume through the membrane, which is inconsistent with previous drift ratchet cases studied. Coupled with the non-reversible circulation effects in the basins, these observations cast doubt on the previous conclusion that particle transport is due to a ratchet mechanism. Klaus et al. (2011) suggest this behaviour may be generic to most pressure driven oscillatory flows, however they state that this finding does not invalidate the ratchet phenomenon, and suggest alternate experimental forcing mechanisms to avoid spurious drift. As these studies (Klaus et al., 2011; Matthias & Muller, 2003) constitute the only reported drift ratchet experiments to date, further work is required to better understand the ratchet mechanism.
These experiments were conducted in the dilute particle regime with concentrations of the order of one particle per pore to simplify elucidation of the drift mechanism. The impact of particle-particle interactions upon performance of the drift ratchet is currently an open question that has not been explored in previous experimental and numerical studies.

Detailed numerical studies of Brenk et al. (2008) and Mehl et al. (2008) use a coupled finite volume scheme to fully resolve the particle-fluid hydrodynamics of a non-diffusive particle within a drift ratchet. These studies ignore Brownian motion and consider significantly larger pressure amplitudes and frequencies than previous simulations (Kettner et al., 2000) and experiments (Klaus et al., 2011; Matthias & Muller, 2003). Whilst these studies accurately resolve the fully coupled particle hydrodynamics in the ratchet, they are highly computationally expensive and well-suited for detailed parametric studies of ratchet phenomena.

Whilst early studies (Brenk et al., 2008; Kettner et al., 2000; Matthias & Muller, 2003; Mehl et al., 2008) did not identify the physical origins of rectified motion in the hydrodynamic drift ratchet, more recent studies have clearly elucidated the governing physical mechanism. Blanchet, Dolbeault, and Kowalczyk (2009) and Kondratyev et al. (2014) studied the dynamics of the Fokker-Planck equation that describes evolution of the particle probability distribution function (PDF), and showed that particle drift can only arises when the asymptotic particle PDF is non-uniform. As such, particle drift is driven by the accumulation of particles in the fluid flow field through hydrodynamic interactions between particles and the pore wall.

For spherical particles under the flow conditions within the ratchet, the particle velocity in the fluid is divergence-free (Schindler, Talkner, Kostur, & Hänggi, 2007) and so particle accumulation cannot occur in this region. Conversely, the particle velocity (defined at the particle centre) near boundaries is not divergence-free due to the geometric requirement that the particle boundary cannot cross the pore boundary. This constraint however is not enforced by steric factors, but rather by the particle-wall hydrodynamic force that diverges as the gap between the particle and pore wall approaches zero. These lubrication forces generate particle accumulation or depletion respectively when the fluid streamlines are converging or diverging with respect to the pore wall. Brownian motion also plays an important role, as it is the interaction of thermal diffusion and particle drift that generates non-uniform particle PDFs; in the absence of diffusion particle trajectories are fully reversible over a forcing period. Hence the physical mechanism for rectified motion in the drift ratchet is persistent particle accumulation that arises from the combination of particle lubrication dynamics and molecular diffusion. It is important to note that reduced particle mobility near pore walls also generates augmented tensorial particle diffusivity (Carbajal-Tinoco et al., 2007; Happel & Brenner, 1983; Perkins & Jones, 1992) that must be quantified to accurately predict particle drift.

Whilst the particle lubrication dynamics for arbitrarily shaped walls is an outstanding fluid mechanical problem, several leading order approximations have been developed (Golshaei & Najafi, 2015; Schindler et al., 2007) in the literature. Golshaei and Najafi (2015) developed an approximation for the particle hydrodynamic mobility tensor via superposition of the particle mobility near a flat wall to mimic the ratchet geometry. This spatially variable mobility is used to approximate the augmented particle diffusivity, and this tensorial diffusivity is found to hinder particle drift to a greater extent than that for the free-space diffusivity. In this study the short range influence of the pore wall on particles is captured via a purely steric interaction which neglects lubrication forces present between the particle and the wall and results in particle reflection from the pore walls. Whilst this approximation appears to qualitatively predict particle drift, the quantitative accuracy of this approximation unclear, and must be established to determine the feasibility of this method to predict particle drift. Due to the difficulty in solving the hydrodynamic lubrication problem for arbitrary shaped walls, accurate quantitative prediction of the drift velocity as a function of the pore geometry, forcing dynamics and particle properties is an open problem. Current reduced transport models, such as Fick-Jacobs, fail to accurately predict the drift velocity, and hence separation characteristics of these microfluidic devices. Therefore accurate prediction of mean field particle dynamics from the pore-scale hydrodynamics is a current hurdle to the development and exploitation of drift ratchets as a viable continuous microfluidic separation technology.

In this paper we perform numerical simulation of a drift ratchet to assess its dynamic similarity. This is to simplify the design and development process of experiments and eventually aid in the design of commercial drift ratchet membranes. In an attempt to further clarify the drift ratchet particle dynamics we have investigated the effect of the spatially varying diffusivity on a hydrodynamic drift ratchet model. The development and validation of numerical drift ratchet model is outlined in Sec. II, and in Sec. III the physical mechanisms that drive the drift ratchet are investigated prior to discussing the results and conclusions in Sec. IV and V, respectively.

# 2. MODEL DEVELOPMENT

To assess dynamic similarity of the drift ratchet and further elucidate the ratchet mechanism we develop a numerical model of particle motion in an infinite drift ratchet. Our model draws inspiration from Golshaei and Najafi (2015) using hard core interactions between finite radius particles and the pore wall, coupled with a spatially varying diffusion coefficient. We combine this with the approximate flow field, calculated using the same method as that from Kettner et al. (2000) as shown in (2). Our numerical model will be used to assess how the drift ratchet mechanism scales.

## 2.1 Particle hydrodynamics

The spatial displacement of a Brownian particle in the bulk of a viscous fluid flow (Burada, Hänggi, Marchesoni, Schmid, & Talkner, 2009) over a time step $\Delta t$ is described by the overdamped Langevin equation,

$$x_{particle}(t) = x_{fluid}(x(t),t) + \sqrt{2D_{th}\Delta t}\,\gamma \qquad (1)$$

where $x_{particle}(t)$ and $x_{fluid}(x(t),t)$ are the displacements of a Brownian particle and the fluid respectively, $D_{th}$ is the free-space particle thermal diffusivity $D_{th} = k_B T / 6\pi\mu R$, $\Delta t$ the time step, and $\gamma$ a Gaussian random variable with unit variance (Kettner et al., 2000). In the absence of fluid flow, numerical solution of (1) over an ensemble of 100 particles recovers the particle mean square displacement associated with the thermal diffusion coefficient. The ensemble statistics were found to be independent of time step for $\Delta t \leq 10^{-6} s$, hence $\Delta t = 10^{-6} s$ was used throughout this study. Assuming dilute particle concentrations and a large number of ratchet pores in series, we neglect both particle-particle interactions and the effect of finite pore length and basins at either end of the ratchet.

The steady fluid velocity field $v_0(x) = \nabla \times \left(\dfrac{\Psi}{r} e_\theta\right)$ in the axisymmetric pore for smooth variations in pore radius $r_p(z)$ is well approximated (Kettner et al., 2000) in terms of the Stokes stream function $\Psi$,

$$\Psi(r,z) = -\frac{1}{2}\left(\frac{r}{r_p(z)}\right)^2 + \frac{1}{4}\left(\frac{r}{r_p(z)}\right)^4 \qquad (2)$$

where $r$, $z$ respectively are the radial and axial coordinates inside the pore. Whilst this expression is only valid for small perturbations of the pore diameter along the pore axis, it serves as a useful approximation to study qualitative behaviour.

The Strouhal number, $St$, is less than $10^{-1}$ for the flow conditions under consideration and as such under transient forcing conditions, the temporal velocity field $v_{fluid}(x(t),t)$ is well approximated by the separable equation

$$v_{fluid}(x(t),t) = v_0(x(t)) g(t) \qquad (3)$$

where $g(t)$ is the forcing protocol. In this study we consider the sinusoidal forcing protocol $g(t) = \sin(2\pi f t)$.

As the particles are neutrally buoyant, have negligible Stokes number and the particle and fluid Reynolds numbers are both negligible, the only mechanism for particles to deviate from fluid trajectories is via particle-wall hydrodynamic interactions and Brownian motion. Other mechanisms such as added mass, buoyancy, lift and Basset forces in the ratchet are also negligible. The impact of particle motion upon the fluid field can also be shown to be negligible, hence we only consider one-way coupling between the fluid and particles, as reflected by (1). Under the approximation of a point particle, the fluid velocity at the centre of the particle was used in the first term of (1), and the rotation of a particle as a result of fluid shear is not considered.

## 2.2 Capturing augmented diffusivity

To study the impact of spatially-variable thermal diffusivity due to reduced particle mobility near the pore wall, we initially perform simulations with a constant diffusion coefficient $D_{th}$ prior to introducing a tensorial diffusion coefficient $D_V(x(t))$. Due to lubrication forces, this diffusivity approaches the free-space diffusivity for large values of the particle-wall gap $h$ and decays to zero as $h$ approaches the particle radius $a$. As the lubrication forces are anisotropic, the resultant particle diffusivity is tensorial. For a particle undergoing diffusion in the presence of an isolated planar wall, the parallel $D_\parallel(h)$ and perpendicular $D_\perp(h)$ components of which are (Happel & Brenner, 1983; Perkins & Jones, 1992)

$$\frac{D_{th}}{D_\parallel(h)} = 1 - \frac{8}{15}\ln(1-\beta) + 0.029 + 0.04973\beta^2 - 0.1249\beta^3 + \ldots \quad (4)$$

$$\frac{D_{th}}{D_\perp(h)} = \frac{4}{3}\sinh\alpha \sum_{n=1}^{\infty} \frac{n(n+1)}{(2n-1)(2n+3)} \times \left[ \frac{2\sinh(2n+1)\alpha + (2n+1)\sinh 2\alpha}{4\sinh^2\left(n+\frac{1}{2}\right)\alpha - (2n+1)^2\sinh^2\alpha} - 1 \right] \quad (5)$$

where $\beta = a/2h$ and $\alpha = \cosh^{-1}(2h/a)$. Both of these relationships have recently been verified experimentally (Carbajal-Tinoco et al., 2007) for colloidal particles diffusing near a planar wall, the problem of diffusion in the presence of curved walls has received little attention and is still an outstanding problem in fluid mechanics. When the lateral and longitudinal curvature of the pore wall are negligible compared to that of the particle, $\frac{1}{r(z)} = \kappa_{lateral} \gg \frac{1}{a}$, $\kappa_{longitudinal} \gg \frac{1}{a}$, the isolated flat wall relationships (4), (5) accurately approximate the particle diffusivity in the ratchet, where $h$ is the smallest particle-wall spacing. The maximum curvature of the wall along the longitudinal direction of the pore is less than the particle curvature, 1.4<2.9. The curvature in the lateral direction follows a similar behaviour, 1.3<2.9. Clearly in our case the curvatures are similar in magnitude and so the validity of this assumption is unclear.

## 2.3 Particle-wall interactions

Rectified particle motion is generated by the combination of thermal diffusion and the hydrodynamic interactions between an advecting particle and the pore wall, as summarised in Sec I. To study the scaling properties of the drift ratchet we simulate these interactions using a simplified model illustrated in Fig. 2.

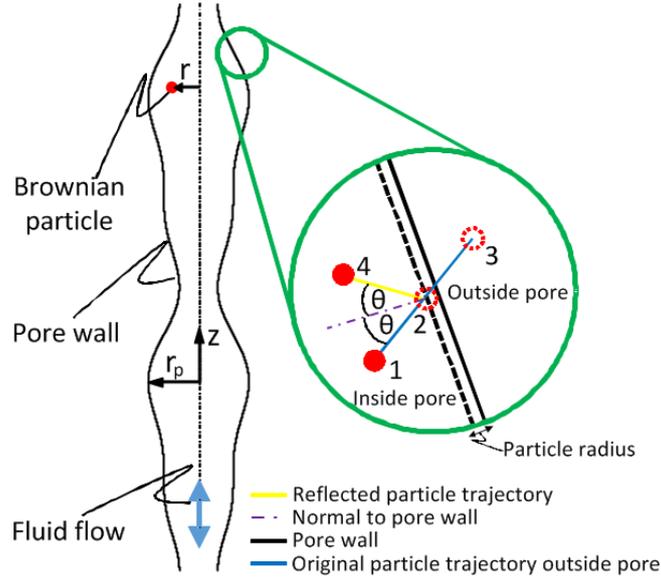

**Fig. 2** Schematic of the particle-wall interactions used in our numerical model. The lengths 2-3 and 2-4 are equal. The geometry of the pore is that studied by Kettner et al. (2000).

Here particles are advected by the fluid velocity and diffuse via Brownian motion as per the overdamped Langevin equation (1). While particle-wall collisions (arising from either advective or diffusive motion) cannot occur for smooth particles as the hydrodynamic resistance diverges logarithmically as the particle-wall gap approaches zero (4), in the same vein as (Kettner et al., 2000), we model the particle-wall hydrodynamic interactions via a reflective boundary condition which qualitatively recovers the same particle clustering behaviour as the complete hydrodynamic interactions. The gross effect of this reflective boundary condition is that it augments the particle PDF near the ratchet walls in a manner that is dependent upon the wall orientation with respect to the fluid streamlines. Specifically, the reflection condition tends to accumulate particles on converging walls and likewise deplete particles on diverging walls, hence the qualitative impact of this condition is similar to that of the true particle-wall hydrodynamic interactions. This reflection condition also recovers the property that the drift velocity decays to zero with decreasing particle size. Whether the reflection condition is quantitatively representative, or can be augmented such that it is, of the full hydrodynamic interaction is currently an open question.

### 2.4 Dimensionless parameters

To develop scaling arguments for the drift ratchet, we define the following dimensionless parameters which are kept constant over the different ratchet sizes: the Péclet number $\text{Pe}$, which captures the relative advection and diffusion timescales, the Strouhal number $\text{St}$, which characterizes the relative viscous and forcing timescales, the ratio of particle to pore size $\alpha$, and the non-dimensional fluid flow amplitude $\beta$.

$$\text{Pe} = \frac{v_{max} d_{min}}{D_{th}}, \tag{6}$$

$$\text{St} = \frac{d_{min}}{T v_{max}}, \tag{7}$$

$$\alpha = \frac{a}{d_{min}}, \tag{8}$$

$$\beta = \frac{T v_{max}}{A}, \tag{9}$$

$v_{max}$ is the maximum fluid velocity within the pore which occurs at the minimum pore diameter $d_{min}$, $T$ is the period of fluid oscillation, $a$ is the diameter of the particle and $A$ is the distance fluid travels along the centreline of the pore over half a period of oscillation.

The remaining dimensionless number is the Reynolds number, $Re$, which is typically less than unity in microfluidics, corresponding to laminar and reversible flow (the maximum Reynolds number in our study was approximately $10^{-2}$). However, it is important to note that fluid recirculation regions can occur within the drift ratchet, even at low Reynolds numbers for certain pore geometries. Such recirculation does not arise for the small smooth undulations of the pore geometry studied herein (Islam, Bradshaw-Hajek, Miklavcic, & White, 2015). Inertial effects associated with acceleration of the oscillating fluid may be considered negligible if the viscous timescale $\tau = \dfrac{d_{min}^2}{\nu}$ ($\approx 10^{-5}$ s) is smaller than the fluid forcing period $T$. This ratio is given by the product of the Reynolds and Strouhal numbers, both of which are small, justifying separability of the temporal velocity field (3). The product $St\,Re$ results in this ratio of viscous timescale to forcing period. The Reynolds number and Strouhal number are small and therefore inertial effects due to the oscillations can be neglected.

## 2.5 Model validation

The parameters used in the drift ratchet simulations are summarised in Table 1. We consider the two cases of ratchet operation investigated by Kettner et al. (2000): in Case 1 the fluid displaced along the centreline of the pore, in half an oscillation period is equal to a single ratchet unit length, whilst under Case 2 the fluid displaced is double the ratchet unit length.

**Table 1** Parameters used in validation of our drift ratchet simulations (Kettner et al., 2000).

| Parameters | Case 1 (1 x amplitude) | Case 2 (2 x amplitude) |
|---|---|---|
| Fluid amplitude ($A$) | 6 μm | 12 μm |
| Flow rate ($Q$) | 2426 μm³s⁻¹ | 4853 μm³s⁻¹ |
| Fluid oscillation frequency ($f$) | 40 Hz | |
| Viscosity ($\mu$) | 0.5 $\mu_{water}$ | |
| $\mu_{water}$ | 1.025×10⁻³ Nsm⁻² | |
| Temperature ($T$) | 293 K | |
| Particle radius ($r$) | 0.35 μm | |
| Boltzmann constant ($k_B$) | 1.38×10⁻²³ m²kgs⁻²K⁻¹ | |
| Minimum pore diameter ($d_{min}$) | 1.52 μm | |
| Reynolds number ($Re$) | Less than 0.008 | |
| Stokes number ($Stk$) | 1×10⁻² - 1×10⁻⁴ | |

To verify the drift ratchet model the calculated average particle drift velocity, $v_e$,

$$v_e = \frac{z(t_{run})}{t_{run}} \tag{10}$$

and effective diffusion coefficient, $D_e$

$$D_e = \frac{\left[z^2(t_{run}) - z(t_{run})^2\right]}{2t_{run}} \tag{11}$$

are compared to those calculated by Kettner et al. (2000), where $z(t_{run})$ is the displacement along the axis of the pore over a time period $t_{run}$ and $\overline{\cdots}$ denotes the ensemble average over 100 particles. Typical motion of an ensemble of particles for the 1x and 2x amplitude cases outlined in Table 1 are illustrated in Fig. 3 below. These results indicate that a doubling of the oscillation amplitude results in reversal of the mean transport direction. The calculated mean drift and diffusion values are shown in Table 2 illustrating that our results compare well with those of Kettner et al. (2000).

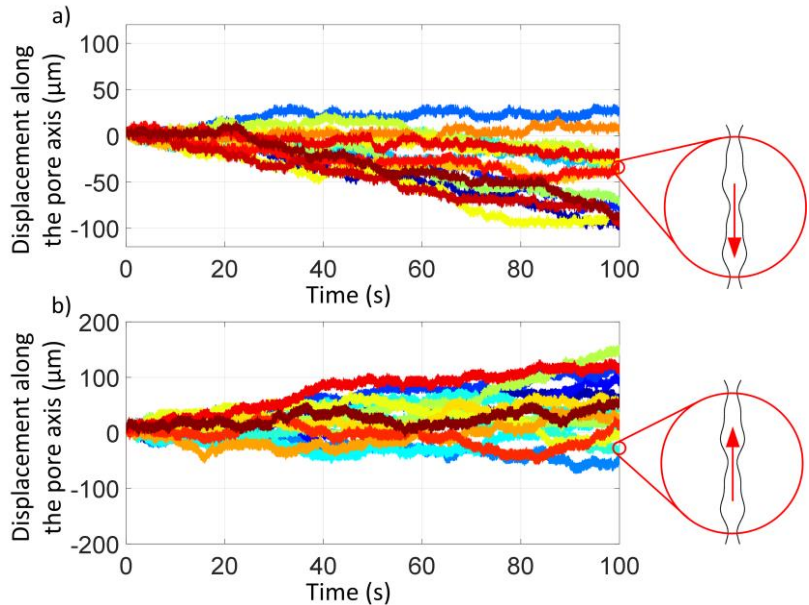

**Fig. 3** Displacement of 15 random particles as a function of time. a) 1x amplitude and b) 2x amplitude as per Table 2.

**Table 2** Comparison of the average drift velocity and effective diffusion coefficient between our model and Kettner et al. (2000) for case 1 and case 2, averaged over 100 particles. A negative drift velocity represents particles moving downwards in Fig. 3.

|  | Case 1 (1 x amplitude) | | Case 2 (2 x amplitude) | |
|---|---|---|---|---|
| Numerical model | Our model | (Kettner et al., 2000) | Our model | (Kettner et al., 2000) |
| $v_e$ (µms$^{-1}$) | -0.41 | -0.46 | 0.39 | 0.45 |
| $D_e / D_{th}$ | 3.12 | 2.45 | 9.6 | – |

The discrepancy in the validation results above could be attributed to the different representations of the particle-wall interactions, and/or the approximation of the fluid flow, or not averaging the fluid velocity over the volume of the particle in our model. The higher effective diffusion coefficient in the 2 x amplitude case, even though it has a smaller or equivalent drift velocity compared to that in the 1 x amplitude case, shows higher variance in total axial displacements over the same time period.

## 3. PARTICLE BEHAVIOR

It has recently been demonstrated (Martens et al., 2013; Schindler et al., 2007) that in the absence of hydrodynamic interactions between the particles and the pore walls, the equilibrium adiabatic particle PDF is uniform across an asymmetric pore, hence the drift velocity is zero. Whilst the particle reflection boundary condition only captures these hydrodynamic interactions in a qualitative sense, the simulations herein recover the limiting hydrodynamic behaviour that particle drift does not occur when the particle radius is zero. This is clearly shown in Fig. 4, where particle displacement is plotted as a function of time for finite diameter particles and point particles. The only difference between a finite radius particle and a point particle is how close the centre of the particle can approach a wall.

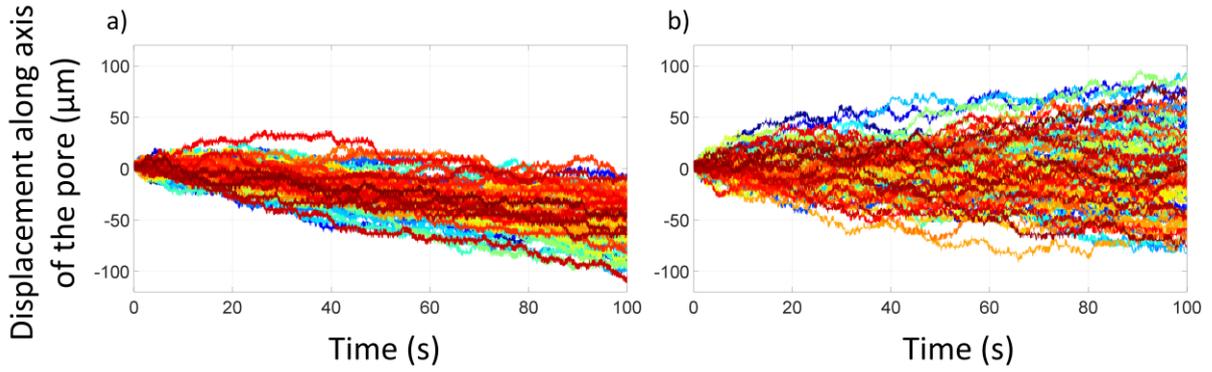

**Fig. 4** Displacement of 100 random particles as a function of time for a non-spatially varying diffusion coefficient. a) Particle-wall interaction with a finite particle radius and b) Particle-wall interaction using point particle.

Point particles (zero radius) in the absence of Brownian motion follow streamlines which cannot intersect the wall. Brownian motion facilitates the traversing of particles across streamlines in an otherwise restrictive laminar flow and move towards the wall. Once finite radius particles are close enough to the wall streamlines can be crossed simply by the hydrodynamic interactions between a finite radius particle and the pore wall. It is this interaction which is necessary to generate particle drift. This concept is illustrated in Fig. 5 that shows the motion of a finite advecting particle near a wall. A particle advecting along streamline A in Fig. 5 is forced onto a path parallel to the pore wall at the edge of the particle exclusion zone (minimum distance from the wall the centre of a particle can occupy due to its finite radius). After travelling through a constriction in the pore, the particle experiences a diverging pore wall and remains on a faster, straighter streamline B in Fig. 5 (Schindler et al., 2007).

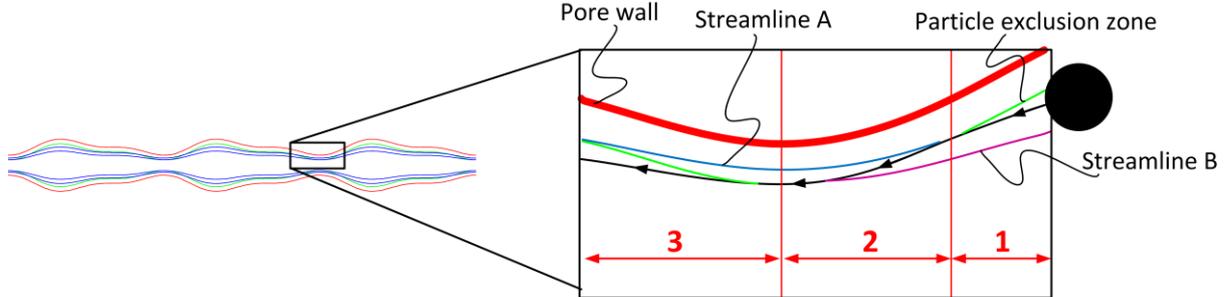

**Fig. 5** Schematic of the mechanism thought to be contributing to driving a drift ratchet adapted from (Schindler et al., 2007).

So how does the reflecting boundary condition affect particle dynamics? To answer this question we examined how the particles are distributed with respect to a pore repeating unit as a function of time. We graphically represent the particle PDF over a periodic ratchet unit in Fig. 6. The particle PDF $\rho(\boldsymbol{x}(t),t)$, averaged over 100 particles, is scaled with the local axial fluid velocity inside the pore to calculate the average drift velocity,

$$g_+(\boldsymbol{x}) = \frac{1}{T}\int_0^{T/2} \rho(\boldsymbol{x}(t),t) v_{fluid}(\boldsymbol{x},t)dt \qquad (12)$$

$$g_-(\boldsymbol{x}) = \frac{1}{T}\int_{T/2}^{T} \rho(\boldsymbol{x}(t),t) v_{fluid}(\boldsymbol{x},t)dt \qquad (13)$$

$$v_e = \sum g_+(x) + \sum g_-(x). \qquad (14)$$

The summation in (14) is over the ratchet unit area shown in Fig. 6.

Maxima of particle probability occur at the edge of the exclusion zone at 0 and 6 μm along the pore as shown in Fig. 6. This is due to the interaction of the particles with the pore wall, moving them to a faster (inner) streamline as previously discussed. Similar to that observed in Schindler et al. (2007) it can be seen in Fig. 6 that particles accumulate on the inside of a converging wall and disperse when the walls diverge.

The particles traverse the width of the pore in the 1x amplitude case as shown in Fig. 6(a) and Fig. 6(c), whereas the radial migration of particles, in the 2x amplitude case, is restricted as depicted in Fig. 6(b) and Fig. 6(d). This restriction comes from the fact that, for the 2x case, no matter where particles are with respect to a ratchet unit the fluid advection term is large enough to make them cross a throat of the pore. This throat wall interaction continually constricts the particle as outlined in Fig. **5**. As we have seen from Kettner et al. (2000) doubling the fluid amplitude can reverse the direction of particle drift. This difference in particle position probability outlined here highlights significant differences between the two cases that can lead to drift reversal.

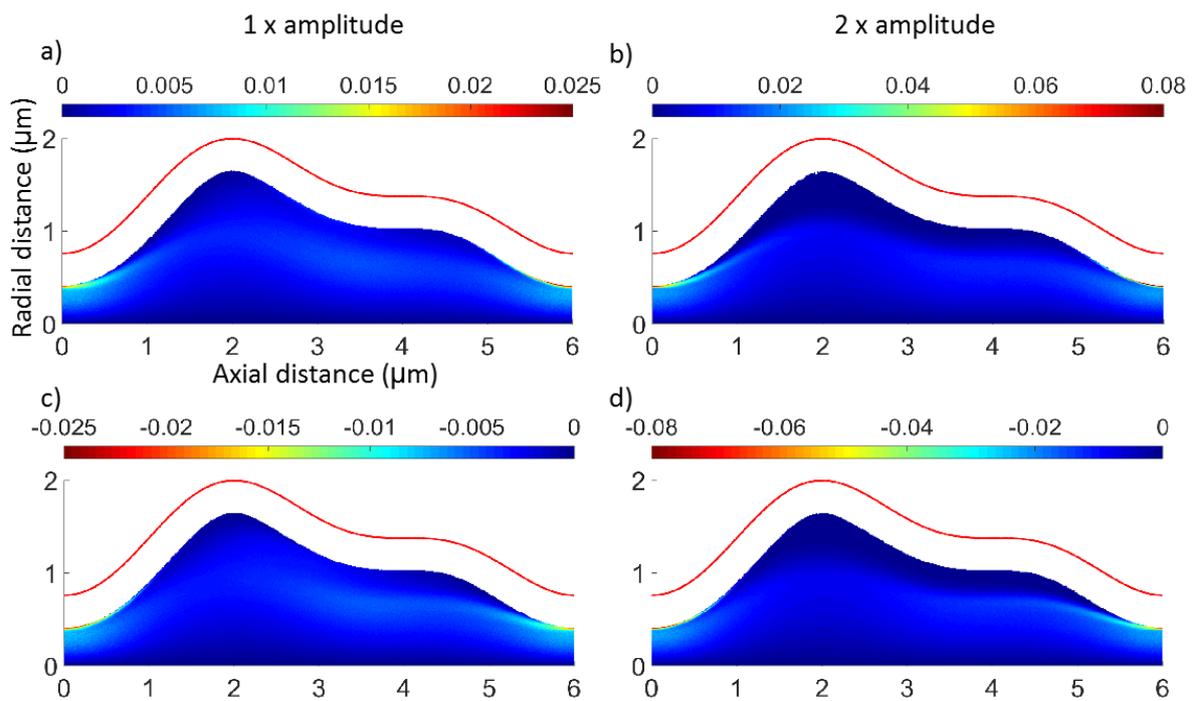

**Fig. 6** The particle probability distribution over a run time of 100 s and 100 particles in a drift ratchet pore for the 1 x and 2 x amplitude case, left and right respectively. a) and b) represent the half of a period of fluid oscillation in the positive direction particles/fluid moving from left to right (12). Whereas, c) and d) is that in the negative direction, particles/fluid moving from right to left (13). The red curve represents the pore wall. The white region between the pore wall and the PDF plot is the particle exclusion zone.

The average drift velocity presented can be recovered from the PDFs illustrated in Fig. 6 and is tabulated in Table 3.

**Table 3** Comparison of average drift velocity from our numerical model and from the PDFs in Fig. 6.

|  | Case 1 (1 x amplitude) | | Case 2 (2 x amplitude) | |
| --- | --- | --- | --- | --- |
|  | Numerical model | PDFs | Numerical model | PDFs |
| $v_e$ (μms$^{-1}$) | -0.41 | -0.41 | 0.39 | 0.28 |

The local Péclet number, calculated with the local fluid velocity, thermal diffusion coefficient and minimum pore diameter, is shown in Fig. 7 for the 1x amplitude case. At a time interval of $10^{-5}$ s either side of the nodes of the sinusoidal wave in Fig. 7, the Péclet number reduces to below unity where diffusion would dominate transport of particles. This a very small percentage of the period of oscillation (0.16%) and thus the particles are dominated by advection in the drift ratchet. In the 2 x amplitude case the percentage of time dominated by diffusion is halved to 0.08% of the oscillation period.

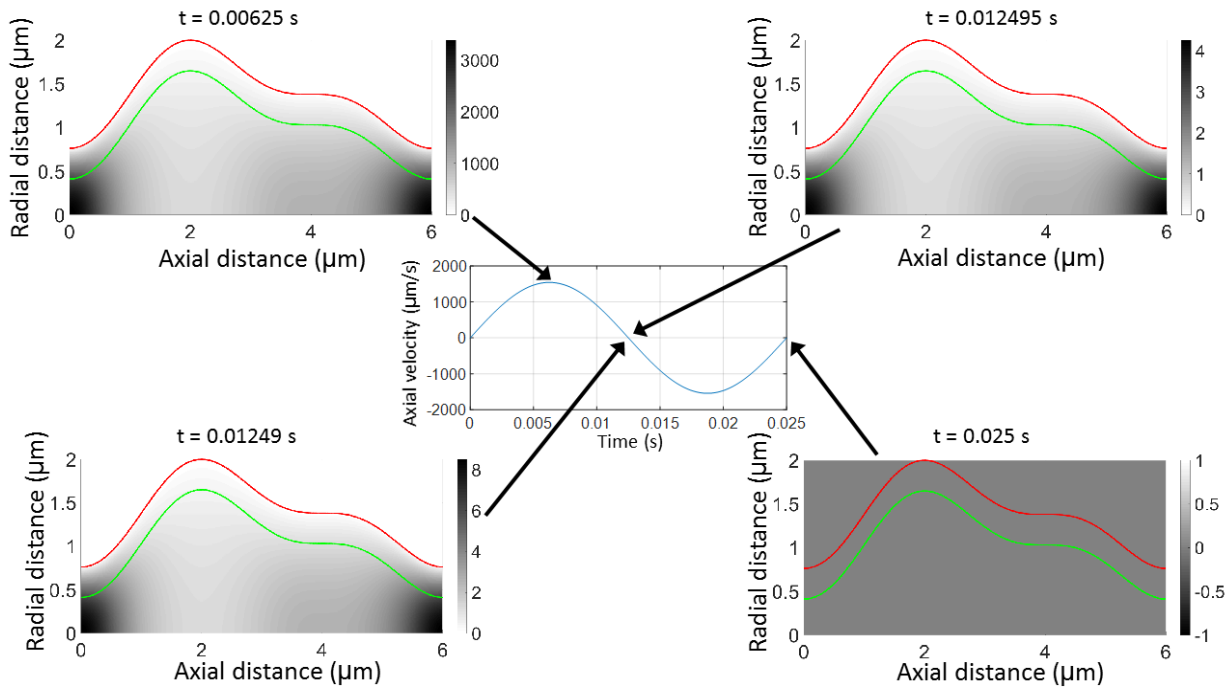

**Fig. 7** Péclet number distribution for a constant diffusion coefficient and length scale based on the minimum pore diameter for the 1 x amplitude case.

## 4. DYNAMIC SIMILARITY ANALYSIS

We have investigated the effect of drift ratchet size on the drift velocity and the effective diffusion coefficient at the various geometric scales relative to the pore size used in Kettner et al. (2000).The shape is the same as that used in the previous section and the cases are outlined in Table 4. Across these cases the Péclet number $Pe$, Strouhal number $St$, the particle/minimum pore diameter ratio $\alpha$ and the dimensionless fluid amplitude $\beta$ are all constant.

**Table 4** Parameters used for the different scaling cases for 1 x amplitude.

| Pore scale | 200% | 150% | 100% | 80% | 60% | 40% | 20% |
|---|---|---|---|---|---|---|---|
| $d_{min}$ (μm) | 3.05 | 2.29 | 1.52 | 1.22 | 0.915 | 0.61 | 0.305 |
| $D_{th}$ (m$^2$s$^{-1}$) | 4.8×10$^{-12}$ | 2.7×10$^{-12}$ | 1.2×10$^{-12}$ | 7.7×10$^{-13}$ | 4.3×10$^{-13}$ | 1.9×10$^{-13}$ | 4.8×10$^{-14}$ |
| $a$ (μm) | 1.4 | 1.05 | 0.7 | 0.56 | 0.42 | 0.28 | 0.14 |
| $L$ (μm) | 12 | 9 | 6 | 4.8 | 3.6 | 2.4 | 1.2 |
| $L^{-1}$ (μm$^{-1}$) | 0.083 | 0.11 | 0.17 | 0.21 | 0.28 | 0.42 | 0.83 |
| $v_{max}$ (μms$^{-1}$) | 5316 | 3988 | 2657 | 2126 | 1595 | 1063 | 532 |
| $Re$ | 3.2×10$^{-2}$ | 1.8×10$^{-2}$ | 7.9×10$^{-3}$ | 5.1×10$^{-3}$ | 2.9×10$^{-3}$ | 1.3×10$^{-3}$ | 3.2×10$^{-4}$ |
| $\alpha$ | 0.46 | 0.46 | 0.46 | 0.46 | 0.46 | 0.46 | 0.46 |

| $Re_p$ | 6.8×10$^{-3}$ | 3.8×10$^{-3}$ | 1.7×10$^{-3}$ | 1.1×10$^{-3}$ | 6.1×10$^{-4}$ | 2.8×10$^{-4}$ | 6.8×10$^{-5}$ |
|---|---|---|---|---|---|---|---|
| $Pe$ | 13329 | 13329 | 13329 | 13329 | 13329 | 13329 | 13329 |
| $St$ | 0.023 | 0.023 | 0.023 | 0.023 | 0.023 | 0.023 | 0.023 |
| $T$ (s) | 0.025 | 0.025 | 0.025 | 0.025 | 0.025 | 0.025 | 0.025 |
| β | 11.1 | 11.1 | 11.1 | 11.1 | 11.1 | 11.1 | 11.1 |

### 4.1 Effect of drift ratchet pore size

The ratio between the effective and thermal diffusion coefficients is constant across different pore sizes, as per Fig. 8. This shows that the ratchet mechanism is independent of pore size if the problem is scaled correctly. That is the relative magnitude of diffusion to advection, as characterized by the Péclet number $Pe$, and relative size of the particle with respect to the pore size $\alpha$, are both held constant. The increased scatter for the higher amplitude case is due to a higher velocity while keeping the time step constant across all the cases. Also included in Fig. 8 are the results of simulations within a straight-walled cylinder to show the effect of an asymmetric pore wall.

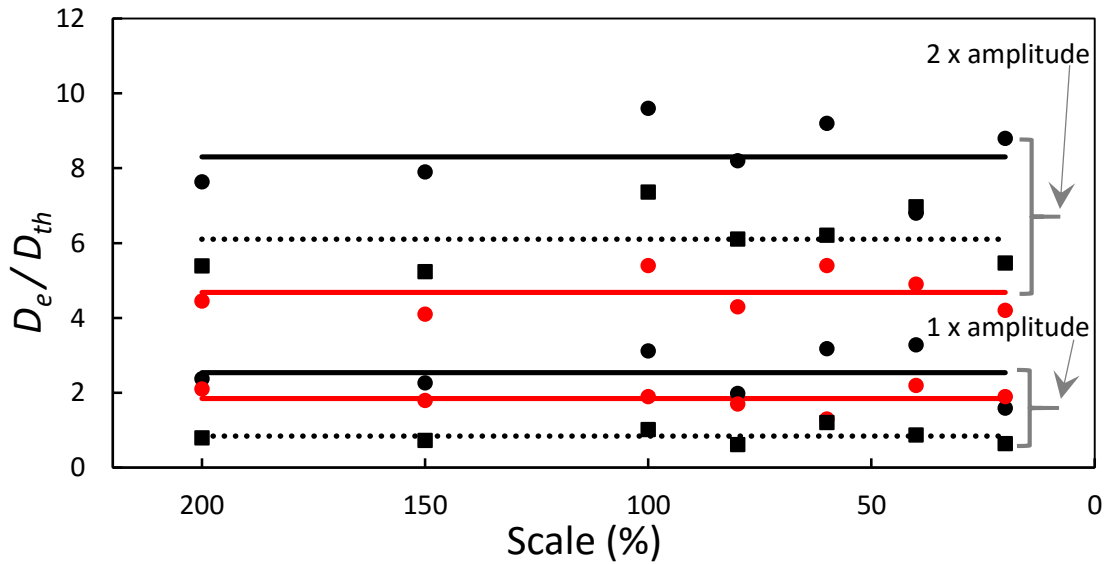

**Fig. 8** Ratio of effective to thermal diffusion coefficients as a function of pore size for $\Delta t = 10^{-6} s$. The circles and solid lines represent simulations with a constant diffusion coefficient, whereas square markers and dotted lines represent spatially varying diffusion coefficient. (Black) Drift ratchet pores and (Red) straight-walled pores.

Whilst one might expect $D_e / D_{th}$ to be unity for a straight walled pore, Taylor-Aris dispersion comes into play, where Brownian particles diffuse longitudinally and radially on similar time-scales. The parabolic shape of the temporally oscillating fluid velocity field affects the effective diffusion coefficient. As expected with plug flow in a straight pore, the longitudinal dispersion is equivalent to the thermal diffusivity as shown in Fig. 9.

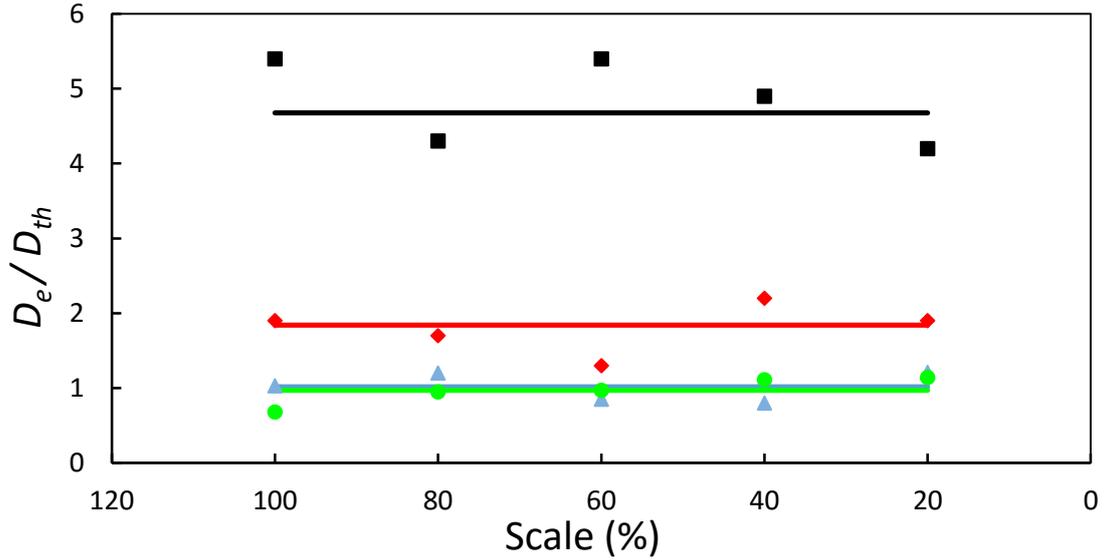

**Fig. 9** Ratio of effective to thermal diffusion coefficients as a function of pore size for straight-walled pores with $\Delta t = 10^{-6} s$. (Black/square) 2 x amplitude with a parabolic velocity profile, (Red/diamond) 1 x amplitude with a parabolic velocity profile, (Blue/triangle) Just diffusion no fluid advection and (Green/circle) 1 x amplitude with a uniform velocity profile.

In order to scale the drift velocity, $v_e$, with pore size we introduce the non-dimensional relative drift velocity,

$$v_{Rel\ drift} = \frac{Tv_e}{L} \qquad (15)$$

where $T$ is the period of fluid oscillation, and $L$ is the axial length of a ratchet period, and so $v_{Rel\ drift}$ is independent of ratchet size as shown in Fig. 10. As expected, the drift velocity for straight walled pores is essentially zero.

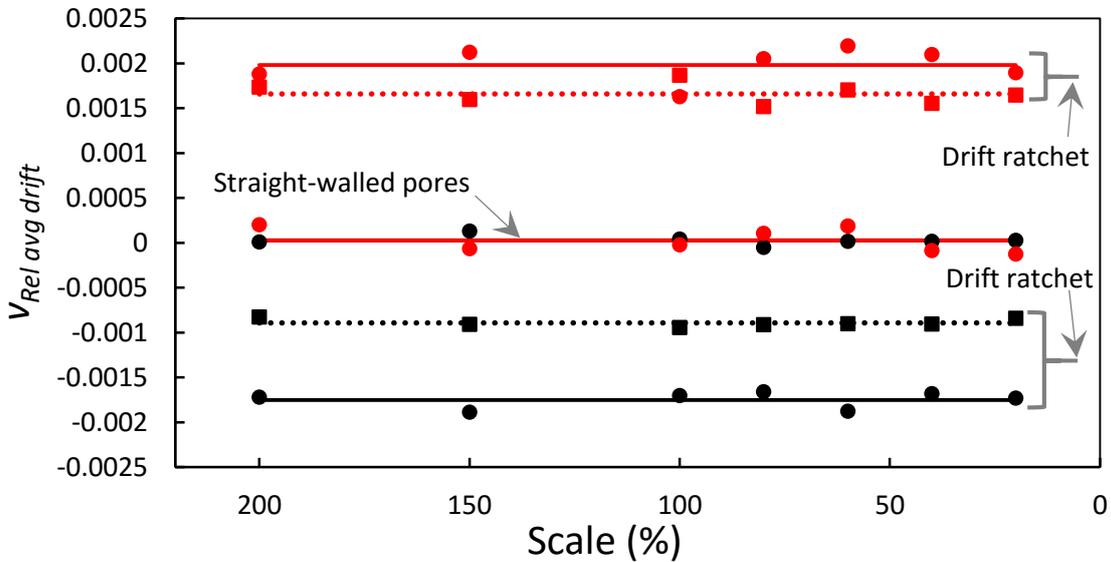

**Fig. 10** Relative drift velocity as a function of pore size for $\Delta t = 10^{-6} s$. The circle markers and solid lines represent simulations with a constant diffusion coefficient whereas square markers and dotted lines represent spatially varying diffusion coefficient. (Black) 1 x amplitude case and (Red) 2 x amplitude case.

## 4.2 Effect of spatially varying diffusion coefficient

As discussed in Sec. II, the thermal diffusion coefficient is both anisotropic and spatially variable near pore walls due to particle-wall hydrodynamic interactions. As shown in Fig. 10 there is only a minor difference between having a constant and a spatially varying diffusion coefficient for the 2x amplitude case, reflecting the fact that diffusion is relatively weak at higher Péclet numbers. This can be explained by understanding that no matter where the particle starts a fluid oscillation cycle with respect to the pore wall it will pass through the throat of the pore. This continuously constrains the particle into the straighter, higher velocity streamlines towards the axis of the pore, where advection dominates diffusion, and the effect of the varying pore diameter is diminished (Motz, Schmid, Hänggi, Reguera, & Rubí, 2014). This mechanism can be observed in the 2x amplitude case in Fig. 6. Conversely, for the 1x amplitude case the drift velocity reduces as to be expected because the diffusion coefficient is monotonically decreasing as it approaches the wall. There is less diffusion and therefore less displacement of the particle in a given amount of time, which leads to a reduction in the ratchet effect. This effect is also apparent in the reduction in the effective diffusion coefficient in both the 1x and 2x amplitude cases. Similar to our results presented herein, Golshaei and Najafi (2015) found that the comparison to a constant diffusivity, a spatially varying diffusivity reduces the particle current through the drift ratchet.

The variation of the constant parameters in the aforementioned plots are shown in **Table 5**.

**Table 5** Variation in effective diffusion coefficient and relative average drift velocity.

|  | Effective diffusion coefficient | | Relative average drift velocity | |
|---|---|---|---|---|
|  | Mean | Relative standard deviation (%) | Mean | Relative standard deviation (%) |
| Drift ratchet 1 x amplitude | 2.5 | ±26.0 | $-1.8\times10^{-3}$ | ±5.3 |
| Drift ratchet 2 x amplitude | 8.3 | ±11.6 | $2.0\times10^{-3}$ | ±9.8 |
| Straight-walled pores 1 x amplitude | 1.8 | ±15.9 | $2.8\times10^{-5}$ | ±190.8 |
| Straight-walled pores 2 x amplitude | 4.7 | ±11.9 | $2.8\times10^{-5}$ | ±479.7 |
| Drift ratchet 1 x amplitude varying diff coefficient | 0.84 | ±25.3 | $-8.9\times10^{-4}$ | ±4.7 |
| Drift ratchet 2 x amplitude varying diff coefficient | 6.1 | ±13.4 | $1.7\times10^{-3}$ | ±7.2 |

## 5. CONCLUSION AND PERSPECTIVE

There is a clear need for more experiments to be completed with respect to drift ratchets to; corroborate the initial results from numerical simulations, prove the existence of the phenomena and to assist in the development of novel applications, fabrication procedures and designs. In terms of simulation work, from our work in this paper highlighting the importance of the short range interactions between a particle and the converging/diverging pore wall, we believe that accurately capturing the particle-wall lubrication dynamics is the next step in achieving a more realistic model. Furthermore, we have shown that if the magnitude of the diffusion, advection and size of a particle is equal relative to the pore scale then the drift ratchet will behave the same across different pore scales. This holds true if the laminar irreversible flow condition is met. This can be used to further develop a reduced model of the drift ratchet phenomenon and more importantly makes it easier to design and compare results from different experiments and develop drift ratchet membranes potentially for commercial use.